\documentclass[%
reprint,
prd,
superscriptaddress,nofootinbib
frontmatterverbose, 
 amsmath,amssymb,
 aps,
 showpacs,
floatfix,
]{revtex4-1}

\usepackage{graphicx}
\usepackage{dcolumn}
\usepackage{bm}
\usepackage{graphicx}
\usepackage{amsmath}
\usepackage{amssymb}
\usepackage{physics}
\usepackage{esvect}
\usepackage{cancel}
\usepackage{bbold}

\usepackage{geometry}
\usepackage{verbatim}
\usepackage{amsmath}
\usepackage{amssymb}
\usepackage{color}
\usepackage{soul}

\usepackage{tikz}
\usetikzlibrary{shapes.misc}
\usetikzlibrary{decorations.markings}

\usepackage[unicode=true,pdfusetitle,
 bookmarks=true,bookmarksnumbered=false,bookmarksopen=false,
 breaklinks=true,
 colorlinks=true,citecolor=steelblue]
 {hyperref}

\makeatletter

\providecommand{\LyX}{\texorpdfstring%
  {L\kern-.1667em\lower.25em\hbox{Y}\kern-.125emX\@}
  {LyX}}

\makeatother

\definecolor{airforceblue}{rgb}{0.36, 0.54, 0.66}
\definecolor{steelblue}{rgb}{0.27, 0.51, 0.71}
\definecolor{amber}{rgb}{1.0, 0.49, 0.0}

\def\be{\begin{equation}}
\def\ee{\end{equation}}
\def\bea{\begin{eqnarray}}
\def\eea{\end{eqnarray}}
\def\bfla{\begin{flalign}}
\def\efla{\end{flalign}}
\def\nn{\nonumber}
\def\pa{\partial}

\def\gsim{\, \rlap{$>$}{\lower 1.1ex\hbox{$\sim$}}\,}
\def\lsim{\, \rlap{$<$}{\lower 1.1ex\hbox{$\sim$}}\,}

\def\hrh{\widehat{r}_h}
\def\hkg{\widehat{k}_g}
\def\hQ{\widehat{Q}}
\def\hl{\widehat{\lambda}}

\definecolor{purple}{rgb}{0.7,0,1}

\begin{document}


\title{Charged black holes in Einsteinian cubic gravity and nonuniqueness}

\author{Antonia~M.~Frassino}
\email[]{antoniam.frassino@icc.ub.edu }
\affiliation{Departament de F{\'\i}sica Qu\`antica i Astrof\'{\i}sica, Institut de Ci\`encies del Cosmos,\\ Universitat de Barcelona, Mart\'{\i} i Franqu\`es 1, E-08028 Barcelona, Spain}

\author{Jorge~V.~Rocha}
\email[]{jorge.miguel.rocha@iscte-iul.pt}
\affiliation{Departamento de Matem\'atica, ISCTE -- Instituto Universit\'ario de Lisboa, Avenida das For\c{c}as Armadas, 1649-026 Lisboa, Portugal}
\affiliation{Departament de F{\'\i}sica Qu\`antica i Astrof\'{\i}sica, Institut de Ci\`encies del Cosmos,\\ Universitat de Barcelona, Mart\'{\i} i Franqu\`es 1, E-08028 Barcelona, Spain}
\affiliation{Centro de Astrof\'{\i}sica e Gravita\c{c}\~ao -- CENTRA, Instituto Superior T\'ecnico -- IST, \\
Universidade de Lisboa - UL,
Av. Rovisco Pais 1, 1049-001 Lisboa, Portugal}

\date{\today}

\begin{abstract}
Black holes are the simplest objects in the universe. They correspond to extreme deformations of spacetime geometry, and can exist even devoid of matter. In general relativity, (electro)vacuum black holes are uniquely determined by their mass, charge and angular momentum.
This feature follows from a uniqueness theorem, which can be evaded if one considers higher dimensions or matter fields coupled to gravity.
Here we find that Einsteinian cubic gravity, a well-motivated modification of Einstein gravity that includes third-order curvature corrections in accordance with low-energy effective theory expectations, admits black hole solutions with charge greater than mass, when minimally coupled to a Maxwell field. 
Moreover, we find that, in this regime, there can be two 
asymptotically flat black holes with the same charge and mass, posing the first example of vacuum black hole nonuniqueness in four dimensions that is free from pathologies. %
Examination of these black hole's thermodynamics reveals that when two branches coexist only the larger black hole is thermodynamically stable, while the smaller branch has negative specific heat.
Einsteinian cubic gravity unveils two further surprising features.
The charged black holes do not possess an inner horizon, in contrast with the usual Reissner-Nordstr\"om spacetime, thus avoiding the need to resort to strong cosmic censorship to uphold determinism.
In addition to black holes, there exists a one-parameter family of naked singularity spacetimes sharing the same mass and charge as the former, but not continuously connected with them. These naked singularities exist in the under-extremal regime, being present even in pure (uncharged) Einsteinian cubic gravity.
\end{abstract}

\maketitle

\section{Introduction \label{sec:Intro}}

A hallmark of the Einstein-Maxwell theory that combines general relativity (GR) with electrodynamics is the validity of the celebrated black hole uniqueness theorem~\cite{Chrusciel:2012jk,Hollands:2012xy,Mazur:2000pn}. It applies to stationary, asymptotically flat, four-dimensional (4D) electrovacuum spacetimes and asserts that all such black hole (BH) spacetimes are determined uniquely by their mass, angular momentum and electric charge.
Of course, the theorem can be evaded either by coupling gravity with different matter or by considering higher dimensions. Concerning the first option, the Bartnik-McKinnon self-gravitating soliton in Einstein-Yang-Mills theory shows that it is sufficient to promote the Maxwell field to a non-Abelian gauge field~\cite{Bartnik:1988am}. Considering  higher dimensions, the discovery by Emparan and Reall of black rings in 5D~\cite{Emparan:2001wn}, when taken in conjunction with the existence of topologically spherical Myers-Perry black holes~\cite{Myers:1986un}, vividly illustrated the inappropriateness of a straightforward application of the theorem. 

A third possibility to evade the uniqueness theorem is to modify the gravitational sector of the theory.
This is the main purpose of the present paper. 
We explicitly demonstrate black hole nonuniqueness in the context of Einsteinian cubic gravity (ECG)---a higher derivative gravity theory that has attracted much attention---when coupled to a Maxwell field.


The nonrenormalizability of GR and the desire to accommodate the observationally inferred cosmological history of the universe within a single framework have sparked a great effort in the past 50 years to devise viable extensions of Einstein gravity. 
Adopting the perspective of low-energy effective theory, the Einstein-Hilbert term in the gravitational action is but the first term in an infinite series of diffeomorphism invariants, with the additional terms being higher order in derivatives and thus becoming increasingly important in the ultraviolet.
Such higher curvature corrections to GR simultaneously enhance the theory's renormalizability behavior~\cite{Stelle:1976gc} and have the potential to accommodate early periods of inflation and late-time acceleration, while avoiding the introduction of dark matter~\cite{DeFelice:2010aj}.
The recently formulated ECG~\cite{Bueno:2016xff} and some of its subsequent extensions incorporate 
all of these appealing features~\cite{Bueno:2016xff,Hennigar:2017ego,Bueno:2017sui,Arciniega:2018fxj,Arciniega:2018tnn}, although its viability as a bare cosmological model has put into question~\cite{Pookkillath:2020iqq}.
In 4D, ECG is the most general diffeomorphism-invariant metric theory of gravity up to cubic order in curvature, whose linearized spectrum on maximally symmetric backgrounds coincides with that of GR, and for which static spherically symmetric vacuum solutions are governed by a single equation of motion.
Moreover, the theory admits a well-defined limit to GR, thus providing a phenomenologically interesting  extension~\cite{Hennigar:2018hza,Poshteh:2018wqy}.

Asymptotically flat black hole solutions of 4D 
ECG were examined in~\cite{Hennigar:2016gkm,Bueno:2016lrh}. The solutions
are described by a single metric function satisfying a nonlinear second-order differential equation [see Eq.~\eqref{eq:EOM} below] that has to be solved numerically, but all thermodynamic quantities can be computed analytically since they follow from a local analysis around the horizon. 
However, some startling aspects went unnoticed, especially regarding nonuniqueness of charged black holes, as well as the existence of positive-energy horizonless solutions. 
We aim to fill this gap here.

A uniqueness theorem in higher derivative gravity was previously obtained in~\cite{Mignemi:1991wa} but it applies only to a restricted class of $f(R)$ theories, for which the Lagrangian is a polynomial in the Ricci scalar, leaving 
ECG (and any metric theory whose Lagrangian contains contractions of multiple Riemann tensors) outside its scope.

The nonuniqueness we uncover is not specific to ECG: it is known to occur in quadratic gravity, even for neutral static BHs~\cite{Lu:2015cqa,Lu:2015psa,Goldstein:2017rxn}. In that case a second branch of BHs exists, in addition to the usual Schwarzschild solution, 
but they show unreasonable pathological behavior. Furthermore, they are necessarily small in Planck units and thus feature large curvatures near the horizon (indicating even higher derivative corrections should be included).
In contrast, we will show that ECG-Maxwell theory with coupling constant above a certain mass-dependent bound contains two competing branches of BHs with the same conserved charges (both of which are regular on and outside the event horizon) 
---as long as the electric charge is greater than the mass. 
In other words, BH nonuniqueness occurs precisely in what would be called the over-extremal regime in Einstein-Maxwell theory.
Finally,  by studying horizonless (but singular) solutions, we find continuously nonunique families of positive-energy naked singularities sharing the same global conserved charges.

\section{Einsteinian cubic gravity coupled to a Maxwell field}
%
In 4D, ECG-Maxwell theory is determined by the action $\mathcal{S}=\int d^4x\, \sqrt{-g} \,{\cal L}$, where the Lagrangian is given by
~\footnote{We consider exclusively the case of vanishing cosmological constant. To be precise, we should also include in this Lagrangian the four-dimensional Euler density already included in Ref.~\cite{Bueno:2016xff}, as well as a further cubic invariant ${\cal C}$ identified in~\cite{Hennigar:2017ego}. However, the former is a topological term, and the latter vanishes identically when evaluated on a spherically symmetric line element of the form we will consider, Eq.~\eqref{eq:metric}. Therefore we omit them for simplicity.}
\be
\mathcal{L} = \frac{1}{16\pi G} \left(R -  2\lambda G^2{\cal P} \right) - \frac{1}{4}F_{ab}F^{ab}\,.
\label{eq:Lagrangian}
\ee
Here, $G$ is the Newton gravitational constant, $R$ represents the Ricci scalar, and the cubic-in-curvature correction to the Einstein-Hilbert action is incorporated in 
\bea
{\cal P} &\equiv& 12 {{{R_a}^c}_b}^d {{{R_c}^e}_d}^f {{{R_e}^a}_f}^b
+ {R_{ab}}^{cd} {R_{cd}}^{ef} {R_{ef}}^{ab} \nn\\
&& -12 R_{abcd} R^{ac} R^{bd}
+8 {R_a}^b {R_b}^c {R_c}^a\,.
\eea
The coupling constant $\lambda$ in Eq.~\eqref{eq:Lagrangian} is chosen to be non-negative; otherwise, the existence of asymptotically flat Schwarzschild-like solutions is precluded~\cite{Hennigar:2018hza} (a more detailed analysis of the case with negative $\lambda$ is discussed in Appendix~\ref{AppA}).
Einstein-Maxwell theory is recovered for $\lambda=0$.
The matter sector is composed only of an Abelian gauge field $A_a$, whose field strength is $F_{ab}=\pa_a A_b - \pa_b A_a$.
From the action above, one derives the Einstein field equations,
\bea
{\cal E}_{ab} &\equiv& E_{acde}{R_b}^{cde} - \frac{1}{2}g_{ab}{\cal L} - 2\nabla^c\nabla^d E_{acdb} \nn\\
&=& 8\pi G \left( F_{ac}{F_b}^c - \frac{1}{4} g_{ab} F_{cd}F^{cd}\right)\,,
\label{eq:EFE}
\eea
where $E^{abcd} \equiv \partial {\cal L}/\partial R_{abcd}$.
These are complemented by the standard Maxwell equations, 
$\nabla_a F^{ab}=0\,$, 
obtained by varying~\eqref{eq:Lagrangian} with respect to the gauge field.

\section{Charged black holes}

Electrically charged spherically symmetric BHs of ECG were studied in~\cite{Bueno:2016lrh}, though not thoroughly. 
When $\lambda=0$
one naturally recovers the Reissner-Nordstr\"om (RN) solution, but at finite (positive) $\lambda$ there can be strikingly marked differences, namely, the absence of a Cauchy horizon in the interior of the BH, the appearance of event horizons in parameter ranges that would naively correspond to over-extremal regimes, and the coexistence of two BHs 
with the same conserved charges under such circumstances, as we now show.

We take the line element to be of the form
\be
ds^2 = g_{ab} dx^a dx^b = -f(r) dt^2 + f(r)^{-1}dr^2 + r^2 d\Omega^2\,.
\label{eq:metric}
\ee
Generic static, spherically symmetric spacetimes need not obey $g_{tt}g_{rr}=-1$ necessarily, but
Eq.~\eqref{eq:EFE} admits solutions with this property~\cite{Hennigar:2016gkm,Bueno:2016lrh}, to which case we shall restrict our considerations.
As for the Maxwell field, we take an electric ansatz, $A_a = A_0(r)\delta_a^t$.
Solving the Maxwell equation yields $A_0(r)=Q/\sqrt{4\pi}r$, where $Q$ is the electric charge of the solution.
Plugging this in the modified Einstein equations~\eqref{eq:EFE} results in a single equation to be satisfied by the blackening factor $f(r)$,
\bea
 2GM &-& \frac{G Q^2}{r} =
-(f-1)r - G^2\lambda \left[ 4f'^3 +12\frac{f'^2}{r} 
\right.\nonumber\\
&-& \left. 
24f(f-1) \frac{f'}{r^2} - 12f f''\left(f'-\frac{2(f-1)}{r}\right) 
\right].\,
\label{eq:EOM}
\eea
The parameter $M$ appears as an integration constant and corresponds to the mass of the spacetime.
This equation is not amenable to exact analytic treatment, so one either resorts to approximations~\cite{Hennigar:2018hza} or to numerical integration.
Here we highlight the most relevant aspects of this procedure.
More details can be found in~\cite{Bueno:2016lrh,Bueno:2017qce,Hennigar:2018hza}.

An analysis of the large-$r$ asymptotic behavior of~\eqref{eq:EOM} reveals that, besides the perturbative corrections (in $\lambda$) to the RN solution $f_{RN}(r)= 1-2GM/r + GQ^2/r^2$, there are also nonperturbative corrections~\cite{Bueno:2016lrh,Hennigar:2018hza}.
While the former are rational functions of $r$ which depend on the charge $Q$, the latter are either growing or decaying exponentials in $r^{5/2}/\sqrt{\lambda}$ (actually modified Bessel functions), which take the same form as in the neutral case. Requiring the absence of the growing mode leaves a three-parameter family of possible asymptotically flat geometries (see Appendix~\ref{AppB} for more details).
On the other hand, roots of $f(r)$ identify possible event horizons of the spacetime, which will be denoted by $r_h$ and happen to be singular points of~\eqref{eq:EOM}.
By Taylor expanding around such a point (assuming $f$ is regular there)
$f(r) = \sum_{n=1}^\infty a_n (r-r_h)^n\,,$
and solving~\eqref{eq:EOM} order by order in powers of $(r-r_h)$, the coefficients $a_n$ can be determined.
The two lowest-order equations form an algebraic system that is used to fix $r_{h}$
and the surface gravity $k_g\equiv f'(r_h)/2 = a_1/2$ in terms of $\lambda$ and $Q$
\footnote{Only those solutions with positive surface gravity are physically relevant. A solution with negative surface gravity in an asymptotically flat spacetime with a continuous blackening factor $f$ can at best correspond to a black hole's inner horizon.},
\begin{subequations}\label{eq:rh_kg_conds}
\bea
&& \hrh - 1 + \frac{\hQ^2}{4\hrh} - 16\hl\, \hkg^2 \left( 2\hkg + \frac{3}{\hrh} \right) = 0\,,
\label{eq:cond1}\\
&& 1 - \frac{\hQ^2}{4\hrh^2} - 2\hkg \hrh - 48\hl \frac{\hkg^2}{\hrh^2} = 0\,.
\label{eq:cond2}
\eea
\end{subequations}
Here we have scaled out the mass $M$ by using dimensionless quantities, defined according to
\bea
\hrh &\equiv& \frac{r_h}{2GM}\,, \;
\hkg \equiv 2GM k_g\,, \;\\
\hQ &\equiv& \frac{Q}{\sqrt{G}M}\,, \;
\hl \equiv \frac{G^2\lambda}{(2GM)^4}\,.\nn
\eea
For $|\hQ|<1$, system~\eqref{eq:rh_kg_conds} has a unique real solution with $k_{g}>0$,
which is to be identified with the black hole's event horizon. 
Here we restrict our attention to positive cubic couplings for the reasons already mentioned.
See Appendix~\ref{AppA} for an analysis of the algebraic conditions~\eqref{eq:rh_kg_conds} when $\lambda<0$.

\begin{figure}
\includegraphics[width=0.40\textwidth]{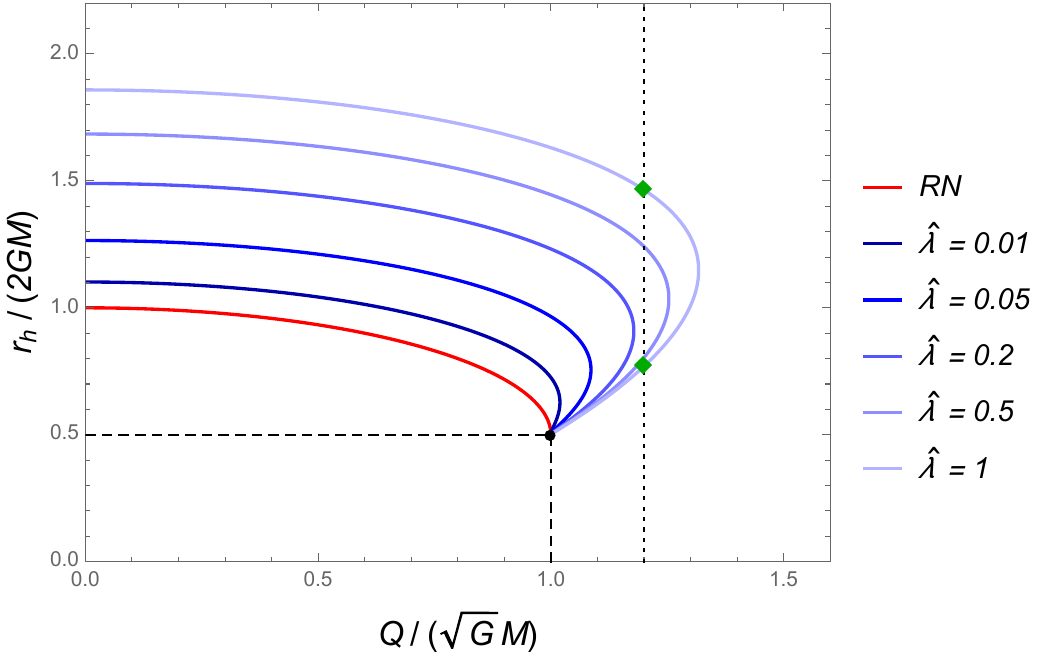}
\caption{
Horizon radius $r_h$ as a function of 
$Q$ (both in units of the mass $M$) for various 
$\lambda$. The lowest red curve corresponds to GR ($\lambda=0$), in which case the black dot at the end of the curve indicates the extremal solution, for which $r_h=GM=\sqrt{G}Q$. Observe that for positive $\lambda$ (actually for $\hl>\widehat{\lambda}_{b}$, more precisely) there are choices of $Q>\sqrt{G}M$ with {\em two} possible solutions for $r_{h}$.
The green diamonds indicate the horizon radius of two coexisting BHs 
with the same $M$ and $Q$, whose profiles are shown 
in Fig.~\ref{fig:f_vs_r}.
\label{fig:rh_vs_Q}
}
\end{figure}

Interestingly for
\be
\label{eq:boundonlambda}
\hl>\widehat{\lambda}_{b} \equiv 1/768
\ee
there exists a finite interval, $1\leq|\hQ|<\hQ_{max}$, where there are {\em two} real solutions with $k_{g}>0$, suggesting that charged BHs in ECG need not comply with the extremality bound $Q\leq \sqrt{G}M$ of Einstein-Maxwell theory. For $|\hQ|>\hQ_{max}$ solutions with $k_{g}>0$ cease to exist. All this is illustrated in Fig.~\ref{fig:rh_vs_Q}. Clearly, $\hQ_{max}$ is a function of $\lambda$, which can be easily determined by finding the root of $dQ/dr_h$.

\begin{figure}
\includegraphics[width=0.45\textwidth]{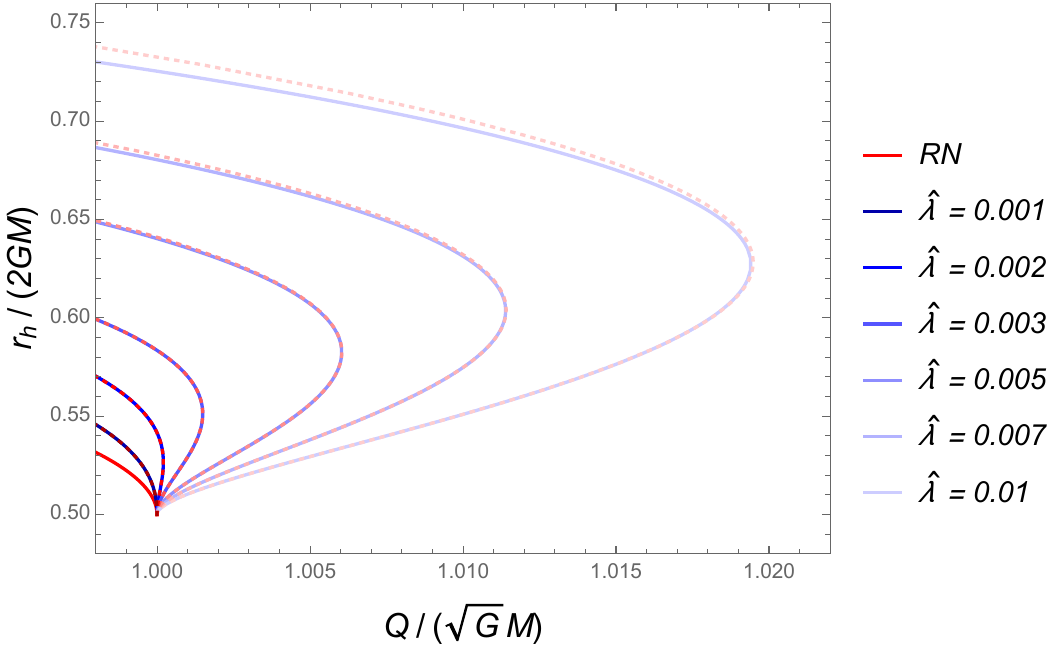}
\caption{Zoom-in around the extremal point (see Fig.~\ref{fig:rh_vs_Q}).
The solid blue lines show the numerical results, while the dotted red curves follow from the perturbative analysis. Different shades correspond to distinct choices of the cubic coupling. There exist over-extremal black holes only when $\hl>768^{-1} \simeq 0.0013$.
\label{fig:rh_vs_Q_zoom}}
\end{figure}

We emphasize that the value $\widehat{\lambda}_{b}$ in Eq.~\eqref{eq:boundonlambda} is exact: it follows from a perturbative study around the extremal point, which is in excellent agreement with the numerics. 
Namely, by perturbing Eq.~\eqref{eq:rh_kg_conds} around the extremal point $(\hrh,\hQ, \hkg)=(\frac{1}{2},1,0)$, using the surface gravity as the small perturbation parameter, one obtains the relation between $\hrh$ and $\hQ$ in parametrized form. For instance, up to cubic corrections in $\hkg$, one finds
\begin{subequations}
\bea
\hrh &=& \frac{1}{2} + \frac{1}{4}\hkg + \frac{1+384\hl}{4}\, \hkg^2 + \mathcal{O}\left(\hkg^3\right)\,,\\
\hQ &=&  1 + \frac{768\hl-1}{8}\, \hkg^2 + \mathcal{O}\left(\hkg^3\right)\,. 
\eea
\end{subequations}
Of course, one can extend the perturbative analysis to arbitrarily high orders in $\hkg$. Including terms up to tenth order is enough to obtain excellent approximations to the numerical results near the extremal point, as illustrated in Fig.~\ref{fig:rh_vs_Q_zoom}.
This shows that charged BHs in ECG can make an excursion to the over-extremal regime, but only when condition~\eqref{eq:boundonlambda} is satisfied.
This bound $\hl > \widehat{\lambda}_{b}$ is mass dependent, so assuming quantum gravity sets in at the Planck scale it translates into a ``uniqueness'' bound, $\lambda\leq G^2 M_{Pl}^4$/48.

%
%
%
%

To confirm that both local solutions for $r_{h}$ are actually associated with asymptotically flat spacetimes, we must integrate~\eqref{eq:EOM} out to large radii and match the corresponding asymptotic behavior.
This is done numerically by determining the coefficients $a_n$ up to some order and then using the truncated series (which has finite radius of convergence) to obtain initial conditions for the integration, starting slightly away from the singular point of the equation $f(r_h)=0$. 
In practice, truncating at $n=10$ is good enough, and an initial integration point $\widehat{r}_i$ displaced by $1\%$ relative to $\hrh$ falls well within the radius of convergence.
The expansion has a single free parameter, $a_2$, in terms of which all other coefficients are fixed. See Appendix~\ref{AppC} for details about the near-horizon expansion.

\begin{figure}
\includegraphics[width=0.38\textwidth]{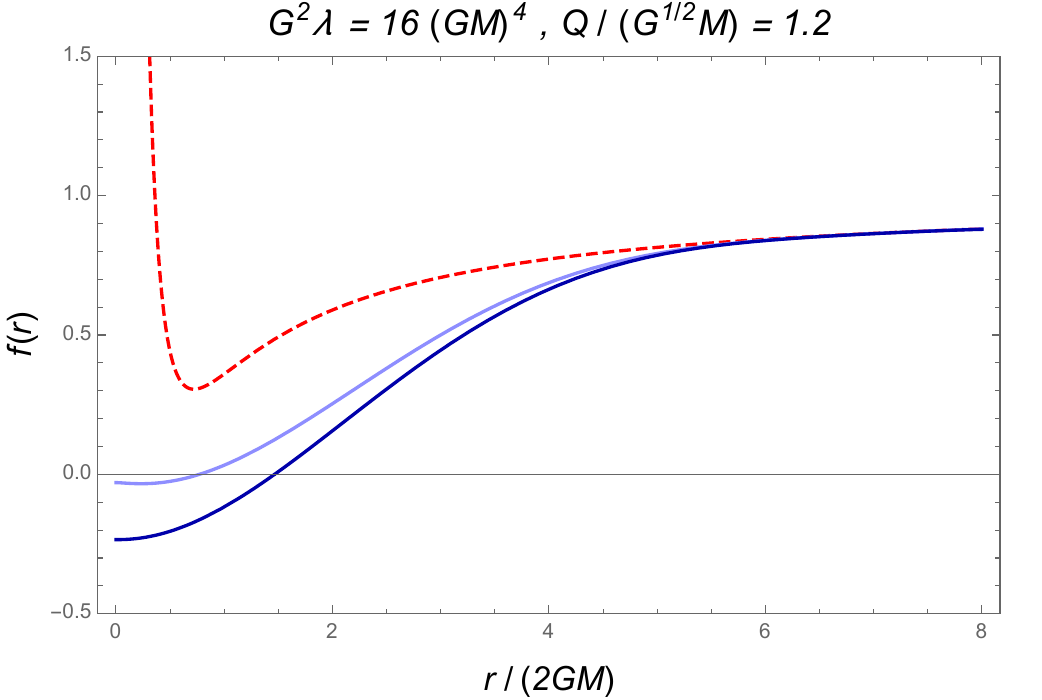}
\caption{
The profile of the blackening factor $f(r)$ for two distinct BH solutions (dark blue 
and light blue solid lines) with the {\em same} conserved charges $M$ and $Q$.
Nonuniqueness is demonstrated for the choices of $\hl=1$ and $\hQ=1.2$
shown in Fig.~\ref{fig:rh_vs_Q} with green diamonds.
The dashed red line is the corresponding RN over-extremal solution, valid for $\lambda=0$.
\label{fig:f_vs_r}
}
\end{figure}

The generic expressions obtained for the expansion coefficients are only valid for nonextremal horizons, $k_g\neq0$.
Indeed, it turns out that the series expansion for the extremal case, $k_g=0$, does not have any free parameter: $a_2$ is also fixed in that particular case.
Focusing on the nonextremal cases, it is always possible to fine-tune $a_{2}$
at the horizon to obtain an asymptotically flat solution that approaches the RN profile asymptotically, even for the cases in which there are two possible values for $r_h$, as shown in Fig.~\ref{fig:f_vs_r}, corresponding to a larger and a smaller BH.
The same behavior of $f(r)$ is reproduced for other values of $\lambda$ and appears to be generic. This entails discrete (twofold) nonuniqueness of BHs in ECG-Maxwell theory. Note the absence of an inner (or Cauchy) horizon in both black hole spacetimes, 
so the causal structure of the interior of these BHs is strikingly different from that of the (subextremal) RN solutions, and instead is qualitatively similar to that of the Schwarzschild solution, thus avoiding any issues with the strong cosmic censorship~\cite{Penrose:1974}.
This behavior appears to be generic for charged BHs in ECG, implying that they do not suffer from any form of mass inflation problem~\cite{Poisson:1989zz}.
In the extremal case, there is no free parameter to adjust and generically a regular asymptotically flat solution cannot be obtained.

\subsection{Thermodynamics of charged ECG black holes}

\begin{figure}
\includegraphics[width=0.41\textwidth]{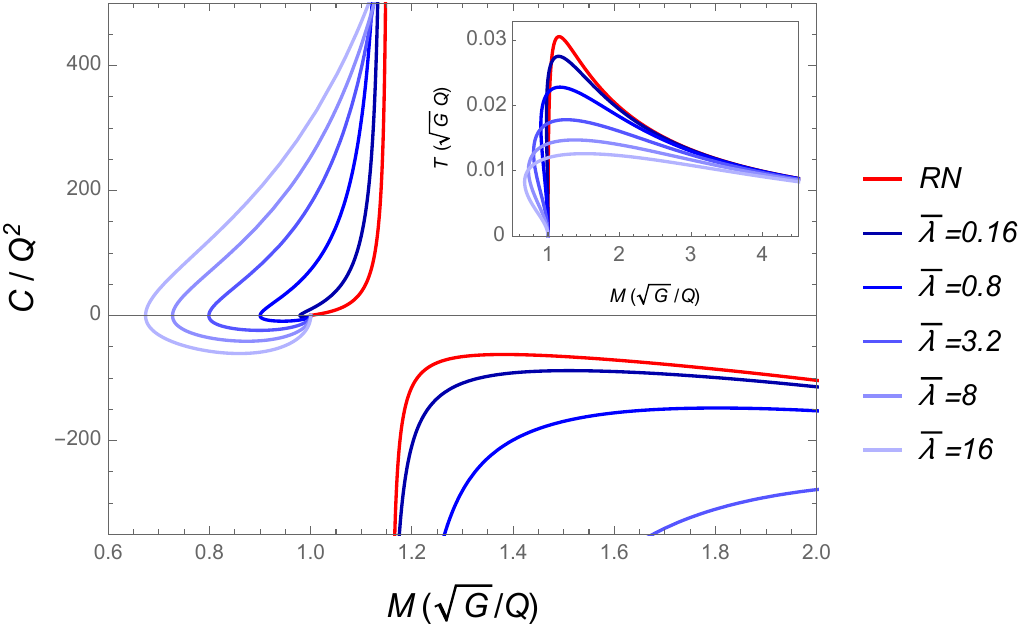}
\caption{
Specific heat as a function of $M$ 
and for different values of $\overline{\lambda}\equiv \lambda/Q^4$, at fixed charge. For $M<Q$ there are two coexisting branches. The larger BHs (which have larger $r_h/M$) display positive specific heat, while smaller BHs have negative specific heat. In the inset, the corresponding temperatures as a function of $M$ (at fixed charge) are shown. 
\label{fig:C_vs_M_wQl}
}
\end{figure}

Similarly to what happens in quadratic gravity~\cite{Lu:2015cqa}, the first law of BH thermodynamics is satisfied by the two coexisting solutions~\cite{Hennigar:2016gkm,Bueno:2016lrh}.
We now check that the larger (smaller) black hole branch has positive (negative) specific heat. Therefore, in the regime of parameters where there is nonuniqueness, only the larger black holes are thermodynamically stable.
By using Wald's formula~\cite{Wald:1993nt,Iyer:1994ys}, it is possible to evaluate the entropy of these black hole solutions~\cite{Bueno:2016lrh}.
The expressions for the temperature $T=k_{g}/2\pi$ and the mass $M=r_0/(2G)$ can instead be obtained from system~\eqref{eq:rh_kg_conds}. 
These are given by
%
%
\bea
    T&=& \frac{r_h^2-\Tilde{q}}{2 \pi  \left(\ell^3+r_h^3\right)}\,,\\
     S&=&\frac{\pi  r_h^2}{G} \left[1-\frac{48 G^2 \lambda  \left(r_{h}^2- \Tilde{q} \right) \left(2 \ell^3- \Tilde{q} r_h+3 r_h^3\right)}{r_h^3 \left(\ell^3+r_h^3\right)^2}\right]\,,\\
     r_0&=&r_h+\frac{\Tilde{q}}{r_h} - \frac{16G^2\lambda(r_h^2-\Tilde{q})^2 \left(5r_h^3-2\Tilde{q}r_h+3\ell^3\right)}{r_h\left(\ell^3+r_h^3\right)^3}\,,\qquad
\eea
where $\Tilde{q} \equiv G Q^2$ and
$\ell \equiv \left[48 G^2 \lambda  \left(r_h^2-\Tilde{q}\right)+r_h^6\right]^{1/6}$.
Useful information regarding the coexisting BHs can be extracted from the behavior of the specific heat at constant charge $C= T \left( \partial S/\partial T \right)_{Q}$ shown in Fig.~\ref{fig:C_vs_M_wQl}. It can be seen that for $\sqrt{G}M<Q$ the two competing BH branches have specific heats with opposite signs.
Working in the canonical ensemble by keeping the charge fixed, it is of interest to evaluate the free energy $F=M-TS$. It turns out that the thermodynamically stable larger BHs are also the ones with lowest free energy~\footnote{There is no sign of critical behavior or first-order phase transitions similar to what was observed for asymptotically anti--de Sitter spacetimes, e.g., for neutral or charged BHs in ECG~\cite{Hennigar:2016gkm,Mir:2019ecg} or for charged BHs in GR~\cite{Chamblin:1999tk}.}.

\section{Naked singularities}

\begin{figure}
\includegraphics[width=0.38\textwidth]{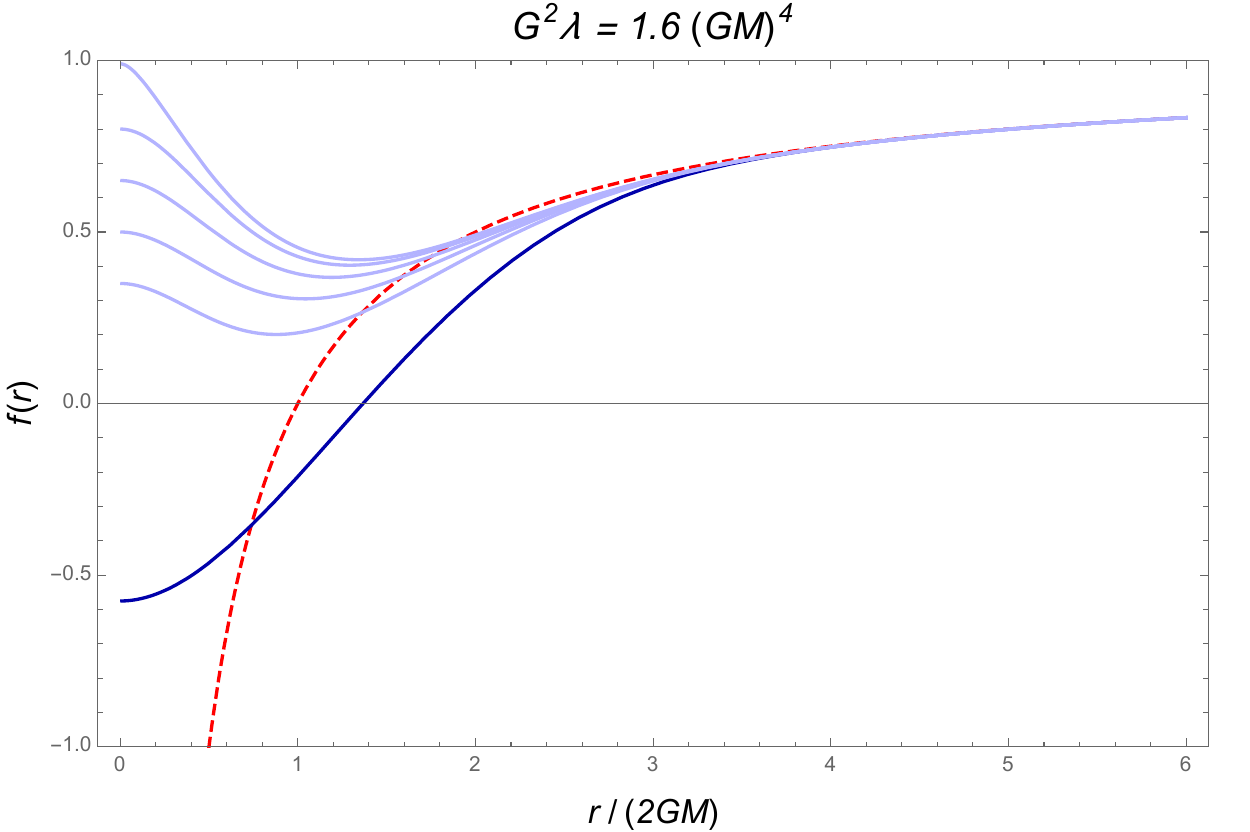}
\caption{
The profile of the blackening factor $f(r)$ for uncharged BHs
(solid dark blue line) and naked singularities (solid light blue lines) in ECG.
All solutions have the same $M$, but they are distinguished by 
different values at the origin.  
The dashed red line represents the Schwarzschild solution 
of GR, which is a valid solution only for $\lambda=0$.
\label{fig:NakedSing}
}
\end{figure}

In addition to the BHs discussed above, the theory admits {\em horizonless} solutions with positive mass but which are nevertheless singular, so they represent naked singularities.
As for BHs, they are determined by a single metric function $f(r)$, the only distinction being that for naked singularities this blackening factor has no roots.
To demonstrate their existence we focus on the neutral case (i.e., vacuum ECG) but similar results are obtained for the charged case.
Uncharged BH solutions can be obtained numerically following the same strategy. 
Consider now integrating, not from the horizon, but from the origin. Equation~\eqref{eq:EOM} also has a singular point at $r=0$. Assuming $f$ is analytic there, it admits an expansion in Taylor series
$f(r) = \sum_{n=0}^\infty c_n r^n\,,$ and the  coefficients of this expansion $c_{n}$ 
can be determined by solving~\eqref{eq:EOM} order by order in powers of $r$. 
The result is that {\em two} of them, namely, $c_0\equiv f(0)$ and $c_2 \equiv \frac{1}{2}f''(0)$, are free parameters, in terms of which all others can be expressed.
In particular, $f'(0)=0$, so that near the origin
\bea
f(r) &=& c_0 + c_2 \, r^2 + \frac{GM}{36 G^2\lambda (1-c_0) c_0} \, r^3 - \frac{1-48 G^2 \lambda c_2^2}{192 G^2 \lambda c_0} \, r^4 \nonumber\\
& & + \frac{GM c_2}{180 G^2 \lambda c_0^2 (1-c_0)} \, r^5 + {\cal O}(r^6)\,.
\eea
Therefore, there is one additional free parameter compared to the expansion around a horizon: 
the mass $M$ and the value of $f$ at the origin, $c_0<1$, (the parameter $c_2$ has to be fine-tuned to get asymptotically flat solutions).
The existence of such horizonless solutions (shown in Fig.~\ref{fig:NakedSing}) does not seem to impose an upper bound on $\hl$, but the range of $c_0$ values that yield asymptotically flat solutions shrinks as $\hl$ decreases, and we were not able to find global solutions for $G^2\lambda \lsim 0.62(GM)^4$~\footnote{It is worth remarking that these solutions can be similarly obtained numerically by assuming the existence of a minimum of $f(r)$ with a positive value, performing the series expansion around such a point (it also possesses two free parameters) and then integrating both outward and inward, instead of starting the integration from the origin.}.
As the Kretschmann scalar near the origin behaves as $\sim 4 (c_0-1)^2 r^{-4}$, the strong curvature region can be made arbitrarily small by fine-tuning $c_0$ close to 1. 
However, it cannot be made fully regular (as was done in~\cite{Holdom:2002xy} for quadratic gravity sourced with incompressible matter) without sending $M$ to zero and retrieving flat spacetime.

\section{Discussion}

We have demonstrated that static charged BHs in 4D ECG differ notoriously from their counterparts in GR in several aspects: nonuniqueness of regular solutions, noncompliance with the extremality bound, and the absence of an inner horizon. Given that the Reissner-Norsdtr\"om solution displays many properties analogous to rotating neutral BHs, it is tempting to speculate that similar features might be present even for vacuum ECG BHs.
Little is known about rotating solutions in ECG.
The only results obtained so far are perturbative in the coupling~\cite{Burger:2019wkq} or refer to the near-horizon geometries of extremal BHs~\cite{Cano:2019ozf}, where similar discrete degeneracy of near-horizon solutions was found.
Interestingly, the existence of ``small'' over-extremal solutions has been related to the weak gravity conjecture~\cite{ArkaniHamed:2006dz}, providing a decay channel for extremal charged BHs.

The horizonless solutions, despite being singular spacetimes, present a more drastic {\em continuous} type of nonuniqueness. Moreover, since the potential $r^{-2}f(r)$ does not feature extrema for these solutions, such naked singularities do not have a photon sphere, so gravitational lensing signatures are markedly 
different from those of BHs~\cite{Hennigar:2018hza,Poshteh:2018wqy}.
An interesting question is whether a naked singularity can be formed under a dynamical process starting from some regular initial state, thus addressing the weak cosmic censorship conjecture~\cite{Penrose:1969pc,Wald:1997wa} in 
ECG.

We have focused on cubic gravity. It remains to be seen if our findings extend to higher derivative gravities, beyond cubic order~\cite{Bueno:2017qce}. 
The effect of adding a cosmological constant or the natural inclusion of cubic terms also in the Maxwell field strength require separate studies.

\begin{acknowledgments}
We thank Roberto Emparan for many useful discussions and Jo\~ao L. Costa, Roberto Emparan and Robert Mann for valuable comments on a draft of the paper.
We acknowledge financial support from the European Union's Horizon 2020 research and innovation programme under ERC Advanced Grant No. GravBHs-692951.
Funding for this work was partially provided by the Spanish MINECO under projects FPA-2016-76005-C2-2-P and MDM-2014-0369 of ICCUB (Unidad de Excelencia ``Mar\'ia de Maeztu'').
J.V.R. thanks FCT for financial support through Project~No.~UIDB/00099/2020.
\end{acknowledgments}


\appendix

\section{Negative coupling constant\label{AppA}}

\begin{figure}
\includegraphics[width=0.45\textwidth]{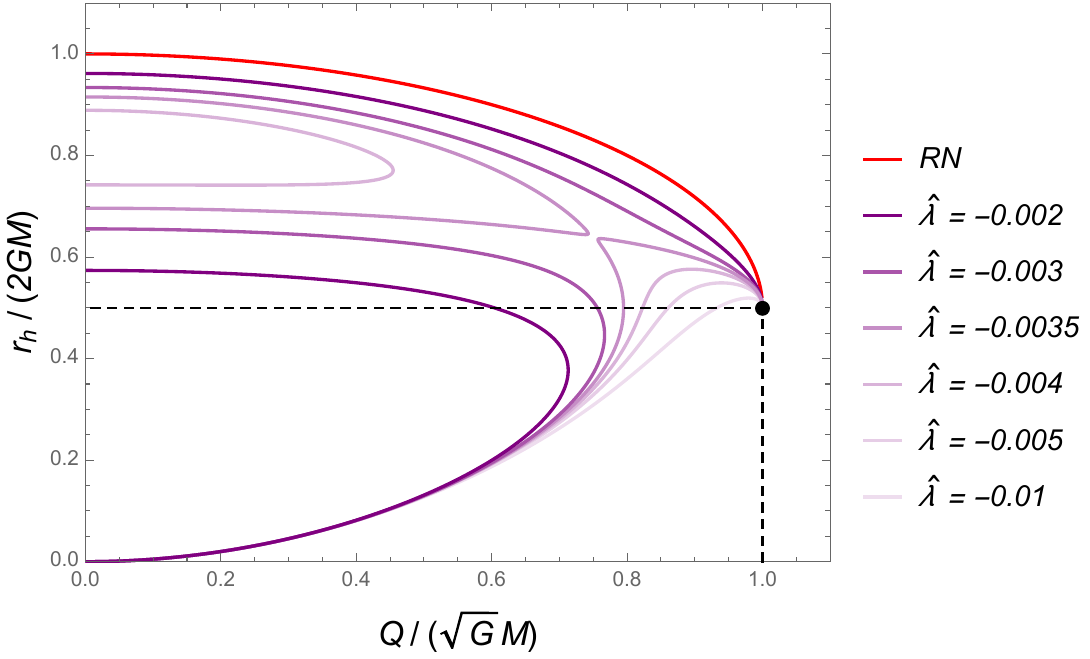}
\caption{Horizon radius $r_h$ ---solutions of~\eqref{eq:rh_kg_conds} with positive temperature--- as a function of the electric charge $Q$ for a selection of nonpositive values of the coupling $\lambda$. The upper red curve corresponds to general relativity ($\lambda=0$). In a certain window of negative values of $\lambda$, there can be up to three possible 
local solutions for the horizon radius. However, neither of them extends to {\em global} asymptotically flat spacetimes that are regular on and outside the horizon.
\label{fig:rh_vs_Q_negLambda}}
\end{figure}

The algebraic system~\eqref{eq:rh_kg_conds} 
\bea
&& \hrh - 1 + \frac{\hQ^2}{4\hrh} - 16\hl\, \hkg^2 \left( 2\hkg + \frac{3}{\hrh} \right) = 0\,,
\label{eq:cond1}\\
&& 1 - \frac{\hQ^2}{4\hrh^2} - 2\hkg \hrh - 48\hl \frac{\hkg^2}{\hrh^2} = 0\,.
\label{eq:cond2}
\eea
admits solutions for the horizon radius (with positive temperature) also when $\lambda<0$. In fact, there can be up to three distinct solutions for $r_h$, as shown in Fig.~\ref{fig:rh_vs_Q_negLambda}. However, these local horizons cannot incorporate
an asymptotically flat spacetime, one that is regular on and outside the horizon. The reason is simple: according to Eq.~\eqref{eq:fasymptotic} below, at large $r$, the blackening factor would become oscillatory instead of approaching the RN behavior. This argument strictly applies only  in the small (negative) coupling regime and is in agreement with numerical explorations. It does not seem to be possible to fine-tune the horizon free parameter in such a way that the integration proceeds to arbitrarily large radius when $\lambda<0$: inevitably a singular point is met, where $f$ diverges irrespective of the magnitude of the coupling constant.

%


\begin{widetext}

\section{Asymptotic expansion \label{AppB}}

Assuming that in the large-$r$ limit the spacetime is 
well approximated by the RN solution plus small corrections, one can determine the asymptotic behavior of $f(r)$, as was done in~\cite{Bueno:2016lrh,Hennigar:2018hza}.
This analysis yields a three-parameter family of solutions of the following form:  
\bea
f(r) &\simeq& 1-\frac{r_0}{r}+\frac{GQ^2}{r^2} - 4G^2\lambda \left\{ 27\frac{r_0^2}{r^6} - 23\frac{r_0^3}{r^7} + \frac{2GQ^2}{r^6} \left[ \frac{2GQ^2}{r^2} \left(19\frac{GQ^2}{r^2}+24\right) - \frac{6r_0}{r} \left(14\frac{GQ^2}{r^2}+9\right) + 57 \frac{r_0^2}{r^2}\right] \right\} \nn\\
&& +{\cal O}\left(\lambda^2,\frac{r_0^3}{r^{11}}\right) + B\, r^{1/4} \exp\left(-\frac{r^{5/2}}{15G\sqrt{\lambda r_0}}\right)\,,
\label{eq:fasymptotic}
\eea
where $r_0 \equiv 2GM$, and the free parameters are the mass $M$, the charge $Q$, and the coefficient $B$.
The general solution (for small $\lambda$) would have an additional term, $A\, r^{1/4} \exp\left(\frac{r^{5/2}}{15G\sqrt{\lambda r_0}}\right)$, but demanding asymptotic flatness imposes $A=0$. On the other hand, the coefficient of the decaying exponential, $B$, has to be fine-tuned so that the asymptotic solution matches smoothly with the near-horizon solution, to which we turn next.

\section{Horizon expansion coefficients\label{AppC}}

In the nonextremal case, $k_g\neq0$, all coefficients in the Taylor expansion of the metric function around the horizon,
\be
f(r) = \sum_{n=1}^\infty a_n (r-r_h)^n\,,
\label{eq:expand_horizon}
\ee
with $n\geq3$ can be expressed in terms of $r_h, k_g, Q$ and $a_2$. For instance, the first two such coefficients are given by
\be
a_3 = \frac{48G^2\lambda \left[ -a_2^2 k_g r_h^3 + \left(3+4k_g r_h\right)\left(a_2 k_g r_h - k_g^2\right)\right] + a_2 r_h^4 + 2 k_g r_h^3 - GQ^2}{144 G^2\lambda\, k_g r_h^2 (1+k_g r_h)}\,,
\ee
\bea
a_4 &=& \frac{1}{55296 G^4\lambda^2 k_g^2 r_h^3 (1+k_g r_h)^2} \left\{ -r_h^5 \left(\frac{G Q^2}{r_h^2}-a_2 r_h^2-2 k_g r_h\right) \right.  \nn\\
&& \!\!\!\!\!\!\!\!\!\!\!\!\!\! - 48 G^2 \lambda \, k_g r_h^2 \left[ \frac{G Q^2}{r_h^2} \left( -\frac{a_2 r_h}{k_g}-7 a_2 r_h^2+9 k_g r_h+4 \right) + k_g r_h \left( \!\left(\frac{a_2 r_h}{k_g}\right)^2 \!+ \frac{a_2 r_h}{k_g}\left(8 a_2 r_h^2-11\right) - 5 a_2 r_h^2-20 k_g r_h-11\right)\right] \nn\\
&& \left. +  
2304 G^4 \lambda^2 k_g^3 \Big[ 5 k_g r_h \left(8 a_2 r_h^2-5\right) + \frac{a_2 r_h}{k_g} \left( a_2 r_h^2 (6 a_2 r_h^2-17)+6 \right) + \left(7 a_2 r_h^2 (5-4a_2 r_h^2) - 6 \right) - 24 k_g^2 r_h^2 \Big] \; \right\}\,.
\eea
Recall the horizon radius and the surface gravity are determined by conditions~\eqref{eq:rh_kg_conds}. Therefore, once the mass $M$, the charge $Q$ and the coupling $\lambda$ are fixed, there is a single free parameter controlling the solutions near the horizon, namely $a_2$.

Considering both the near-horizon and the asymptotic regions, the total number of free parameters (two) is just the right amount to allow at  most a discrete set of smooth solutions to the second-order differential equation~\eqref{eq:EOM}.
%
%
Clearly, the extremal case ($k_g=0$) has to be treated separately. Under this condition, the solution of~\eqref{eq:rh_kg_conds} is simply $r_h = GM$ and $Q = \pm \sqrt{G}M$ and the first nontrivial coefficients of the series~\eqref{eq:expand_horizon} read
\bea
a_2 &=& \frac{1}{(GM)^2}\,, \\
a_3 &=& -\frac{2}{(GM)^3} \frac{ (GM)^4+16 G^2\lambda }{(GM)^4 - 48G^2 \lambda}\,, \\
a_4 &=& \frac{3 (GM)^{12} + 128 G^2 \lambda (GM)^8 - 2304 G^4 \lambda^2 (GM)^4 + 73728 G^6\lambda^3}{(GM)^4 \left[(GM)^4-144 G^2\lambda \right] \left[(GM)^4-48G^2\lambda \right]^2}\,.
\eea
Hence, an extremal horizon has no free parameter to adjust once the mass and the coupling are fixed. A regular, asymptotically flat solution is not expected to exist in this case.


\end{widetext}
\bibliography{solitons_in_ECG_refs}

\end{document}